\documentclass[prl,twocolumn]{revtex4}
\usepackage{hyperref}
\usepackage{amssymb, amsmath}
\usepackage[dvips]{graphicx}
\usepackage{latexsym}
\usepackage{verbatim}

\begin{document}

\title{Transport measurements across a tunable potential barrier in graphene}%
\author{B.~Huard}
\author{J.A.~Sulpizio}
\author{N.~Stander}
\author{K.~Todd}
\author{B.~Yang}
\author{D.~Goldhaber-Gordon}
\email[Corresponding author~: ]{goldhaber-gordon@stanford.edu}
\affiliation{Stanford University, Department of Physics, Stanford, California, USA}%

\begin{abstract}\
The peculiar nature of electron scattering in graphene is among many
exciting theoretical predictions for the physical properties of this
material. To investigate electron scattering properties in a
graphene plane, we have created a gate-tunable potential barrier
within a single-layer graphene sheet. We report measurements of
electrical transport across this structure as the tunable barrier
potential is swept through a range of heights. When the barrier is
sufficiently strong to form a bipolar junctions (npn or pnp) within
the graphene sheet, the resistance across the barrier sharply
increases. We compare these results to predictions for both
diffusive and ballistic transport, as the barrier rises on a length
scale comparable to the mean free path. Finally, we show how a
magnetic field modifies transport across the barrier.
\end{abstract}
\maketitle

The recent discovery by Novoselov \emph{et al.}~\cite{Novoselov2004}
of a new two-dimensional carbon material -- graphene -- has
triggered an intense research
effort~\cite{Zhang2005,Berger2004,Geim2007}. Carriers in graphene
have two unusual characteristics: they exist on two equivalent but
separate sublattices, and they have a nearly linear dispersion
relation. Together these ingredients should give rise to remarkable
transport properties including an unusual suppression of
backscattering~\cite{Ando1998}. Unlike in conventional metals and
semiconductors, electrons in graphene normally incident on a
potential barrier should be perfectly transmitted, by analogy to the
Klein paradox of relativistic quantum
mechanics~\cite{Katsnelson2006,Pereira2006,Cheianov2006}.
Backscattering would require either breaking of the ``pseudospin''
symmetry between electrons living on the two atomic sublattices of
the graphene sheet~\cite{Cheianov2006} or a momentum transfer of
order the Fermi wavevector $k_F$, which can only be produced by a
sharp, atomic-scale step in the potential. In addition to the
obvious relevance to future graphene-based electronics,
understanding transport across potential barriers in graphene is
essential for explaining transport in graphene close to zero average
density, where local potential fluctuations produce puddles of n-
and p-type carriers~\cite{Yacoby2007}. We report an experiment in
which a tunable potential barrier has been fabricated in graphene,
and we present measurements of the resistance across the barrier as
a function of the barrier height and the bulk carrier density.

\begin{figure}[hptb]
\begin{center}
\includegraphics[width=8.5cm]{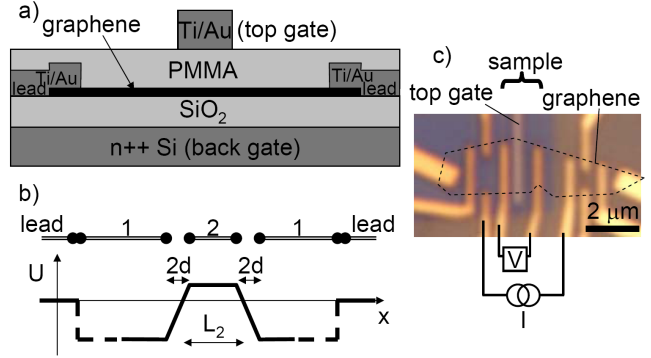}
\caption{a) cross-section view of the top gate device. b) simplified
model for the electrochemical potential $U$ of electrons in graphene
along the cross-section of a). The potential is shifted in region 2
by the top gate voltage and shifted in both regions 1 and 2 by the
back gate voltage. c) Optical image of the device. The barely
visible graphene is outlined with a dashed line and the PMMA layer
appears as a blue shadow. A schematic of the four-terminal
measurement setup used throughout the paper is shown. \label{Fig1}}
\end{center}
\end{figure}

To create a tunable potential barrier in graphene, we implemented a
design with two electrostatic gates, a global back gate and a local
top gate (Fig.~\ref{Fig1}a.) A voltage $V_b$ applied to the back
gate tunes the carrier density in the bulk of the graphene sheet,
whereas a voltage $V_t$ applied to the top gate tunes the density
only in the narrow strip below the gate. These gates define two
areas in the graphene sheet whose densities $n_2$ -- underneath the
top gate -- and $n_1$ -- everywhere else -- can be controlled
independently (Fig.~\ref{Fig1}b.) The graphene was deposited by
successive mechanical exfoliation of natural graphite crystals using
an adhesive tape (Nitto Denko Corp.)~\cite{Novoselov2004}.
 After
exfoliation, thin graphite flakes were transferred onto a chip with
280~nm thermal oxide on top of an n++ Si substrate, used as the back
gate. A single-layer graphene sheet was identified using an optical
microscope, and 30~nm thick Ti/Au leads were evaporated onto the
sheet using standard e-beam lithography. To form the dielectric
layer for the top gate, a thin layer of PMMA (molecular mass 950K,
2\% in anisole) was then spun onto the Si chip at 4000~rpm for 35~s
and baked at $180^{\circ}\mathrm{C}$ for 2 minutes. On one section
of the graphene sheet, the PMMA was cross-linked by exposure to
30~keV electrons at a dose of $21,000~\mu\textrm{C
cm}^{-2}$~\cite{Zailer1996}. The unexposed PMMA was removed by
soaking the chip for 10 minutes in acetone. Finally, a 50~nm thick,
300~nm wide Ti/Au strip was deposited on top of the cross-linked
PMMA using standard e-beam lithography to form the top gate
(Fig.~\ref{Fig1}c.) Raman spectroscopy of the sheet indicates that
it is in fact composed of a single
layer~\cite{Ferrari2006,Graf2007,EPAPS}.

\begin{figure}[hbtp]
\begin{center}
\begin{minipage}{.49\linewidth}
\includegraphics[height=4.2cm]{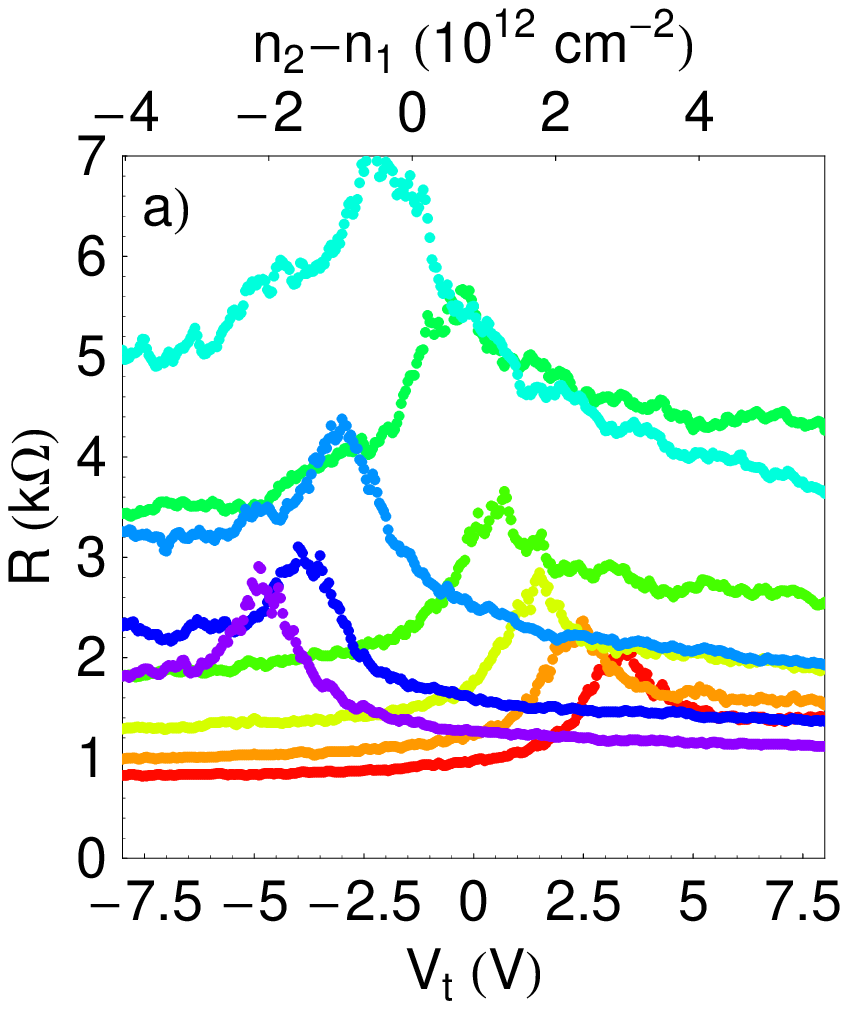}
\end{minipage} \hfill
\begin{minipage}{.49\linewidth}
\includegraphics[height=4.2cm]{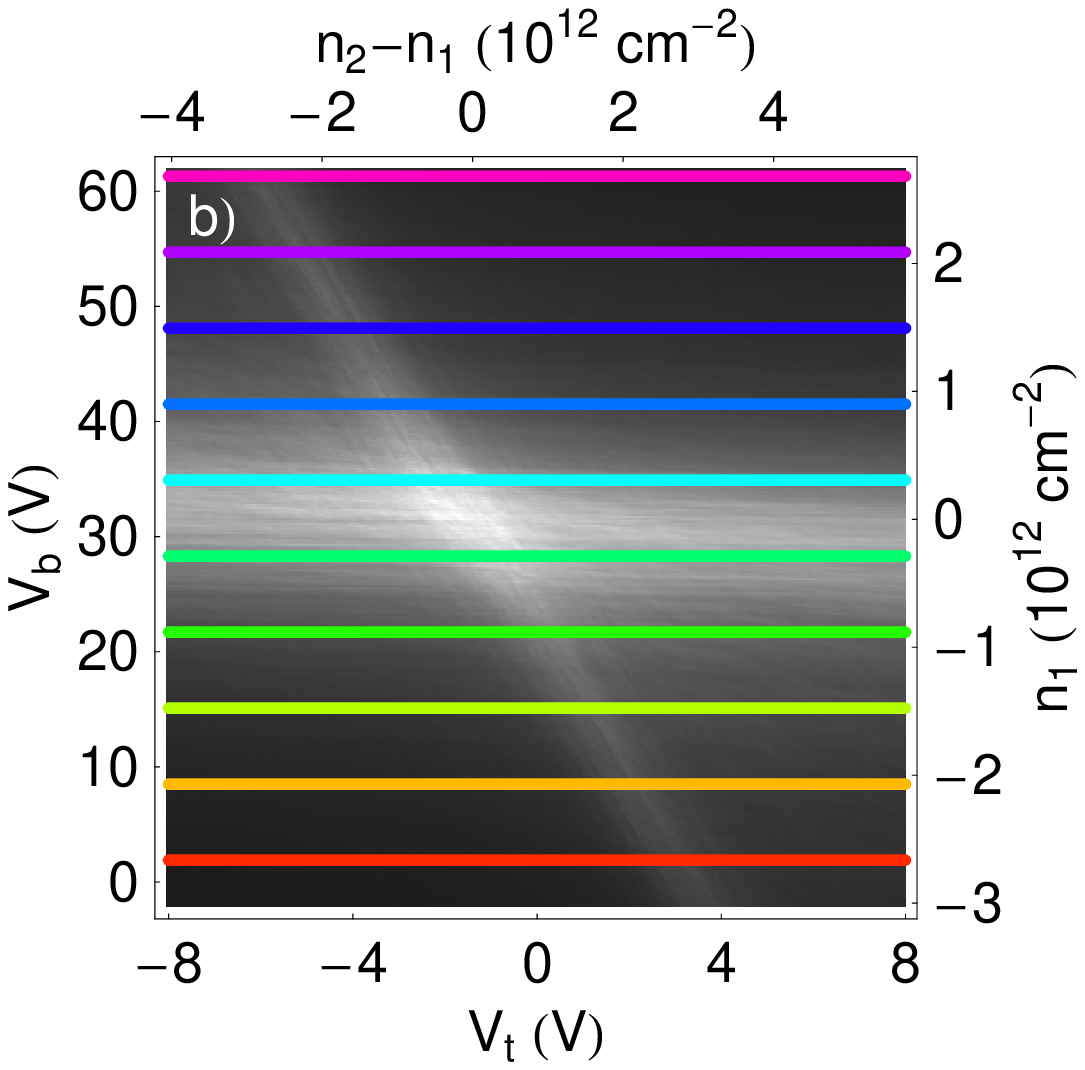}
\end{minipage}
\caption{a) Resistance across the graphene sample at 4~K as a
function of the top gate voltage for several back gate voltages,
each denoted by a different color. b) Two-dimensional greyscale plot
of the same resistance as a function of both gate voltages. Traces
in a) are cuts along the correspondingly-colored lines.\label{Fig2}}
\end{center}
\end{figure}

We performed conventional four-terminal current-biased lockin
measurements (100~nA at 12.5~Hz) of the resistance of a PMMA-covered
$L=1.3~\mu\mathrm{m}$ by $w=1.7~\mu\mathrm{m}$ section of a graphene
flake, in which a potential barrier can be induced by the top gate
(Fig.~\ref{Fig1}c). All the measurements were performed in liquid
Helium at 4~K. In Fig.~\ref{Fig2}b, this resistance is plotted as a
function of both the back gate voltage $V_b$ and the top gate
voltage $V_t$ \footnote{The leakage currents of back and top gates
are linear in voltage over the full range studied, with leakage
resistances $R_\mathrm{b}\approx 3.7~\mathrm{G}\Omega$ and
$R_\mathrm{t}\approx 1.7~\mathrm{G}\Omega$}. Two clear white lines
appear, indicating local maxima in the resistance in the different
regions of the graphene sheet. In the bulk of the sheet (region 1),
the average carrier density is given by
$n_1=C_b(V_b-V_b^\mathrm{0})/e$, where the back gate capacitance per
area $C_b\approx 14~\mathrm{nF~cm}^{-2}$ is inferred from Hall
measurements of other graphene flakes on the same oxide layer. The
horizontal white line marks the neutrality point ($n_1 = 0$) in this
region, allowing us to estimate $V_b^\mathrm{0}\approx
31.5~\mathrm{V}$. In region 2, the average density
$n_2=n_1+C_t(V_t-V_t^\mathrm{0})/e$ is modulated not only by the
back gate, but also by the top gate, with capacitive coupling $C_t$.
The voltage $V_t^\mathrm{0}$ is not necessarily zero, since the
chemical doping in regions 1 and 2 can differ, and this difference
can vary widely from sample to sample. The diagonal white line marks
the neutrality point underneath the top gate ($n_2 =0$). The slope
of this line provides the relative coupling of the graphene sheet to
the two gates: $C_t/C_g \approx 6.8$, so $C_t\approx 1.0\times
10^2~\mathrm{nF.cm}^{-2}$. Using the PMMA thickness of 40~nm as
measured by AFM, this value leads to a dielectric constant
$\epsilon_{\mathrm{PMMA}}=4.5$ (close to the accepted room
temperature value for non-cross-linked PMMA). The crossing point of
these white lines yields
$V_t^\mathrm{0}=-1.4~\mathrm{V}$~\cite{EPAPS}.

Transport fluctuations seen in Fig.~\ref{Fig2} are a reproducible
function of gate voltages and magnetic field (universal conductance
fluctuations~\cite{Rycerz2006}) with amplitude $\delta G\approx 0.2
e^2/h$~\footnote{This value is measured by sweeping magnetic field
at 17 values of the density $n_1=n_2$ between $-3.0\times 10^{12}$
and $-2.3\times 10^{12}\mathrm{cm}^{-2}$.}. The magnetoresistance is
almost perfectly symmetric in magnetic field, as
expected~\cite{EPAPS}. We extract the phase coherence length
$L_\varphi\approx 4~\mu\mathrm{m}$ and the inter-valley scattering
length $L_\mathrm{i-v}\approx 0.15~\mu\mathrm{m}$ from the weak
localization peak. $L_\varphi> L_\mathrm{i-v}$ indicates that the
sample is lying flat on the
substrate~\cite{EPAPS,McCann2006,Morozov2006}.

\begin{figure}[hbtp]
\begin{center}
 \begin{minipage}{.49\linewidth}
\includegraphics[height=4.2cm]{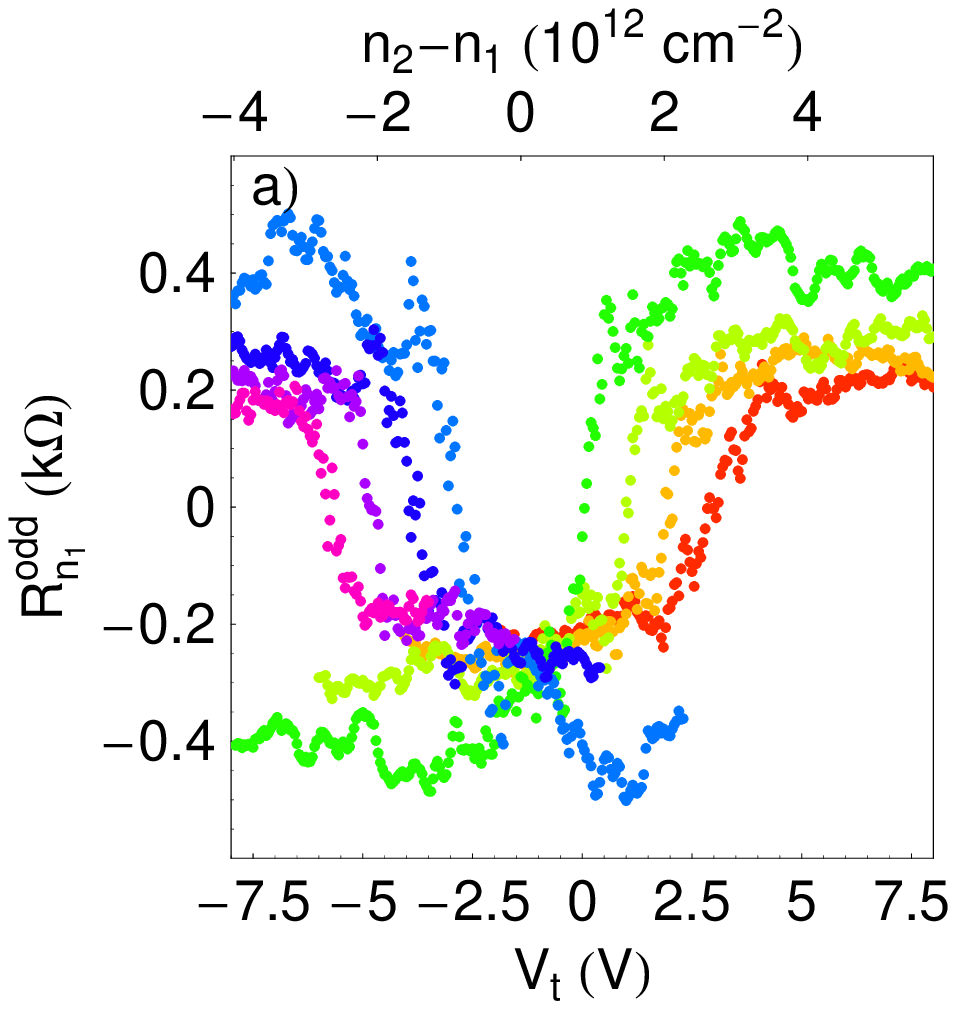}
 \end{minipage} \hfill
 \begin{minipage}{.49\linewidth}
\includegraphics[height=4.2cm]{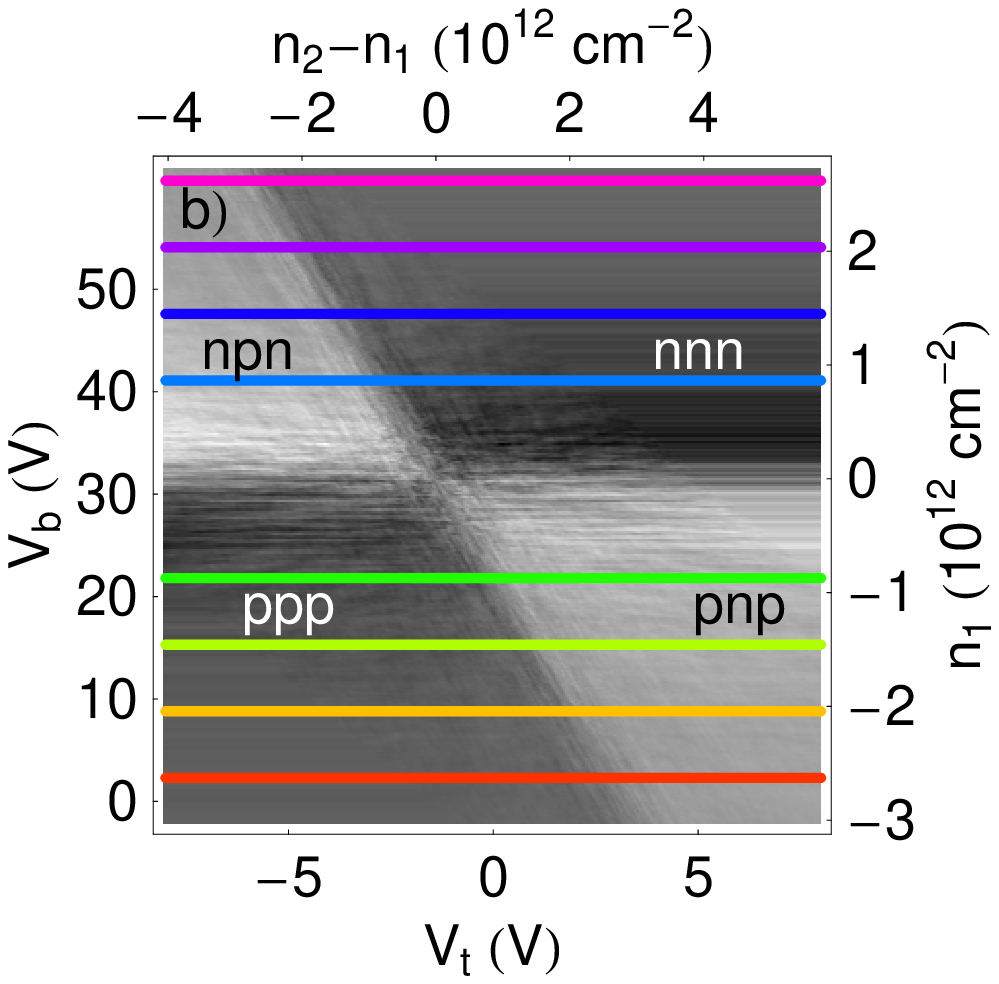}
\end{minipage}
\caption{a) Odd component of the resistance: the part which depends
on the sign of the density $n_2$ in region 2. b) Greyscale plot of
the odd component of the resistance for many values of $n_1$.
Colored lines are the cuts shown in a). \label{Fig3}}
\end{center}
\end{figure}

Fig.~\ref{Fig2}a shows selected cuts through Fig.~\ref{Fig2}b:
device resistance across the potential barrier as a function of the
top gate voltage $V_t$, for several values of the bulk density
$n_1$. Each curve has a maximum close to $n_2=0$, arising from the
enhanced resistivity of the graphene in region 2. However, each
curve is noticeably asymmetric with respect to this maximum: the
resistance depends on whether or not the carriers in region 2
(electrons or holes) are the same type as those in region 1.
Specifically, for given absolute values of the densities $n_1$ and
$n_2$, the resistance is always higher if the carrier types in the
two regions are opposite ($n_1 n_2<0$) than if the carriers are the
same throughout. In order to highlight the effects of pn junctions
between regions 1 and 2, we extract for each value of $n_1$ the part
of the resistance $R^{\rm odd}$ which depends on the sign of the
carriers in region 2: $R_{n_1}(n_2) = R_{n_1}^{\rm even}(n_2) +
R_{n_1}^{\rm odd}(n_2)$, $R_{n_1}^{\rm even}(-n_2) = R_{n_1}^{\rm
even}(n_2)$ and $R_{n_1}^{\rm odd}(-n_2) = -R_{n_1}^{\rm odd}(n_2)$
(Fig.~\ref{Fig3}) \footnote{The voltage $V_t^{(n_2=0)}$ at which the
density $n_2$ is zero is determined using the above equation
$n_2=n_1+C_t(V_t-V_t^\mathrm{0})/e$~\cite{EPAPS}.}. The resistance
of the device away from the junctions (region 1 and possibly the
interior of region 2) does not depend on the sign of $n_2$ and hence
is entirely contained in $R^{\rm even}$, which we do not examine
further.

The presence of pn junctions between regions 1 and 2 is associated
with a substantial increase of the overall resistance (white areas
compared to black areas of Fig.~\ref{Fig3}b). We label each section
of Fig.~\ref{Fig3}b with the carrier types in region 1 and 2, ``p''
for holes and ``n'' for electrons, to emphasize that enhanced
resistance is associated with an npn or pnp junction. The
junction-sensitive resistance curves are almost symmetric upon
simultaneous sign change of both densities $n_1$ and $n_2$:
left-right reflection and color swap in Fig.~\ref{Fig3}a, or 180
degree rotation about the center in Fig.~\ref{Fig3}b. Deviations
from this symmetry are presumably associated with uncontrolled
spatial fluctuations in the density. Strikingly, a sharp step
appears in the resistance as the boundary between nnn and npn or pnp
configurations is crossed. We can explore the underlying physics of
such a resistance increase in two opposite regimes: strongly
diffusive or ballistic. Which applies to experiments depends on
whether the elastic mean free path is greater than or less than the
length over which the potential barrier rises.

\begin{figure}[hptb]
\begin{center}
\includegraphics[width=8.5cm]{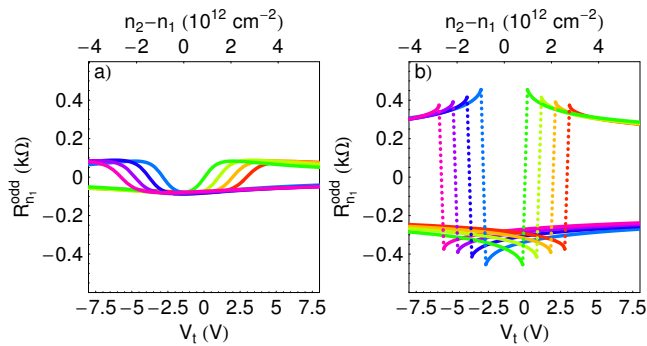}
\caption{a) Using the strongly diffusive model described in the
text, one can predict the resistance as a function of the top gate
voltage $V_t$ for several values of the density $n_1$ (each
represented by the same color as in Fig.~\ref{Fig3}). Here we plot
the odd part of this calculated resistance (cf. Fig.~\ref{Fig3}) for
a barrier smoothness $d=40~\mathrm{nm}$ and a width
$w=1.7~\mu\mathrm{m}$. b) Similar curves for several densities $n_1$
using the ballistic model described in the text. The curves are
plotted only for densities $n_2$ not too close to zero:
$n_2>10^{11}~\mathrm{cm}^{-2}$ (the model diverges at $n_2=0$) and
dashed lines link the curves at opposite sides of $n_2 = 0$.
\label{Fig4}}
\end{center}
\end{figure}

In a simple model, the electrostatic potential $U(x)$ induced by the
set of two gates in the graphene sheet changes linearly between
regions 1 and 2 over a width $2d$ (Fig.~\ref{Fig1}b). Assuming
perfect screening in the graphene sheet (a good approximation for
our relevant density range), we estimate $d \approx 40~\mathrm{nm}$
independent of carrier density, by solving the Laplace potential in
the region between the top gate and the graphene. If the elastic
mean free path $l_e$ is much shorter than $d$, the total resistance
can be estimated by integrating the local resistivity across the
device. The conductivity $\sigma(x)$ at any position $x$ depends
only on the density of charge carriers $n(x)$ and can be inferred
from resistance measurements where the density is uniform
\cite{EPAPS}. It can be approximated by an interpolation between low
and high density behavior $\sigma(x)\approx\left[\left(e\mu
n(x)\right)^2+\sigma_\mathrm{min}^2\right]^{1/2}$~\cite{Geim2007}
where the mobility is $\mu\approx 2\times
10^3\mathrm{cm}^2\mathrm{V}^{-1}\mathrm{s}^{-1}$ and
$\sigma_\mathrm{min}\approx 4e^2/h$. Using $d=40~\mathrm{nm}$ and a
width $w=1.7~\mu\mathrm{m}$, we plot the predicted odd part of
$R(V_t)$ with respect to the voltage $V_t^{(n_2=0)}$ for each value
of $n_1$ on Fig.~\ref{Fig4}a.

At high enough density, where the conductivity $\sigma$ of graphene
is proportional to the density of charges, one can define a mean
free path $l_e$ in an isotropic diffusive model by
$\sigma=2k_Fl_ee^2/h$, where $k_F=\sqrt{\pi n}$ is the effective
Fermi wavevector defined relative to the K-point of the Brillouin
zone~\footnote{Notice that the prefactor 2, which is related to the
inverse of the dimension of the diffusive material in this model,
might be a slight underestimate because of the angle-dependent
scattering in graphene.}. For $n_1=n_2\approx
-10^{12}~\mathrm{cm}^{-2}$, one can see in Fig.~\ref{Fig2}a that
$\sigma\approx 6 \frac{2 e^2}{h}$~\footnote{The conductivity can be
estimated for $n_2=n_1$ as $\sigma=\frac{L}{R w}$ where $R$ is the
resistance.}. Therefore, $l_e\approx 0.03~\mu\mathrm{m}$ for
$n\approx -1\times 10^{12}~\mathrm{cm}^{-2}$. The mean free path is
then already of the order of $d$ for a few volts applied to the back
gate. Hence, the diffusive model is a poor approximation and is
unable to reproduce the data shown in Fig.~\ref{Fig3}.

The opposite regime of ballistic transport has been considered in
the limits of sharp and smooth potential steps. We use here the
results of Cheianov \emph{et al.} \cite{Cheianov2006}, which are
valid in the limit $|n_1|,|n_2|\gg d^{-2}$. Note that the results
for a sharp barrier \cite{Katsnelson2006} lead to the same
qualitative results for $R_{n_1}^{\rm odd}(n_2)$ but are more than
10 times smaller than what follows. The transmission probability
$\tau(\theta_1)$ for electrons impinging from region 1 on the
interface between regions 1 and 2 depends strongly on the angle of
incidence $\theta_1$ with respect to normal incidence, and is given
by
\begin{equation}
 \tau(\theta_1) = e^{-2\pi^{3/2} d
\frac{|n_1|}{|n_1|^{1/2}+|n_2|^{1/2}} \sin^2\theta_1
}\hspace{0.5cm}\mathrm{ if }~n_1n_2<0.\label{Eq1}
\end{equation}

We incorporate both interfaces of the potential barrier into the
resistance calculation in the following way. In the fully ballistic
regime ($l_e\gg L_2,d$, where $L_2$ is the distance between the two
interfaces), all possible reflections at the interfaces, and their
interferences, must be taken into account. However, in the regime of
this experiment, where $l_e \sim d\ll L_2$, the two interfaces can
be considered as independent resistors in series. Using a Landauer
picture, the (antisymmetric) conductance through the full barrier is
then
\begin{equation}
G_\mathrm{npn}=\frac{1}{2}\frac{4e^2}{h}\sum_{m_y=-m_y^\mathrm{max}}^{m_y^\mathrm{max}}
e^{-2\pi^{1/2} d\frac{(2\pi m_y/w)^2}{|n_1|^{1/2}+|n_2|^{1/2}}},
\end{equation}
where we sum the transmission coefficients of incident modes having
all possible values of the quantized wavevector component
perpendicular to the interface $k_y=2\pi m_y/w$, with $m_y$ integer
and
$m_y^\mathrm{max}=\left\lfloor\sqrt{\pi\mathrm{min}(|n_1|,|n_2|)}w/2\pi\right\rfloor$.
Each such mode carries a conductance $4e^2/h$, where the degeneracy
factor 4 is characteristic for graphene. The presence of two
interfaces in series gives an overall prefactor of $1/2$.

Using $d=40~\mathrm{nm}$ and $w=1.7~\mu\mathrm{m}$, we plot the
calculated odd part of the resistance in our ballistic model as a
function of $V_t$ in Fig.~\ref{Fig4}b for the same densities $n_1$
as in the strongly diffusive case, assuming that the resistance in
the case where the sign of $n$ is uniform is much smaller than
$1/G_\mathrm{npn}$. Although the results of the ballistic model are
of the same order of magnitude as the measured values of the odd
part of the resistance, the model is unable to reproduce some
aspects of the experimental data. In particular, the model does not
explain why $R_{n_1}^\mathrm{odd}$ does not only jump but continues
to increase as $|n_2|$ passes beyond zero density. This behavior is
particularly surprising as, for fixed $n_2$, $R_{n_1}^\mathrm{odd}$
{\em decreases} with increasing $|n_1|$. Furthermore, by
construction, the model does not apply close to zero density $n_1$
or $n_2$.

A complementary test of the unusual transmission through a potential
barrier in graphene as a function of the angle of incidence was
proposed by Cheianov \emph{et al.}~\cite{Cheianov2006}. In the fully
ballistic regime, the resistance across the barrier is predicted to
increase as soon as a magnetic field applied perpendicular to the
graphene sheet gets higher than a typical value $B_\star\approx
\hbar/e\left(\sqrt{\pi |n_2|}/\pi L_2^2d\right)^{1/2}$, where $L_2$
is the barrier length ($0.3~\mu\mathrm{m}$ via SEM). Using the
values $d=40~\mathrm{nm}$, $L_2=0.3~\mu\mathrm{m}$, and $n_2=2\times
10^{12}\mathrm{cm}^{-2}$, one gets $B_\star\approx 0.1~\mathrm{T}$.

\begin{figure}[hptb]
\begin{center}
\includegraphics[width=6cm]{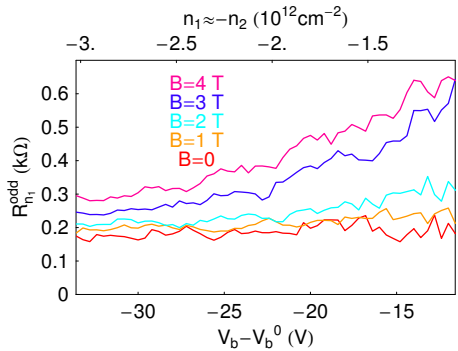}
\caption{Odd part of the resistance as a function of the back gate
voltage $V_b$, with the constraint $n_2 = -n_1$. Measurements were
taken at 4~K for several values of the perpendicular magnetic field
$B$.\label{Fig5}}
\end{center}
\end{figure}

Fig.~\ref{Fig5} shows measurements of the antisymmetrized resistance
across the device for several values of a magnetic field $B$ applied
perpendicular to the graphene sheet, with the top gate voltage $V_t$
adjusted to maintain $n_2\approx -n_1$. This odd part of the
resistance shows an enhancement as the magnetic field increases
above $2.5~\mathrm{T}$, an order of magnitude higher than the
predicted transition field $B_\star$. The fully ballistic model is
therefore unable to explain the results of our measurements in
magnetic field. In fact, the diffusive model also predicts an
increase of $R_{n_1}^{\rm odd}$ as $B$ increases, due to
magnetoresistance of the strip near zero density between p and n
regions. If the entire graphene sheet is set to zero density, its
resistance increases with magnetic field similarly to $R_{n_1}^{\rm
odd}$ at $n_2\approx -n_1\approx 2\times 10^{-12}\mathrm{cm}^{-2}$
(Fig.~\ref{Fig5}). This qualitative agreement indicates that while
ballistic physics plays a role in transport across the individual pn
junctions, transport through the full npn barrier is far from
ballistic.

In conclusion, we have fabricated a gate-tunable barrier device from
a single-layer graphene sheet, and have hence created bipolar
junctions within graphene. We study transport across the potential
barrier as a function of Fermi level and barrier height, as well as
in the presence of an external magnetic field, and demonstrate that
a sharp increase in the resistance occurs as the potential crosses
the Fermi level. This increase is better described by a ballistic
than a diffusive model, but the dependence of the additional
resistance as a function of the barrier height is not yet
understood. A clear explanation of this behavior is essential for
realizing many exciting proposed graphene electronics applications,
such as electron focusing~\cite{Cheianov2007}.

We thank V.I. Fal'ko, E.A. Kim, D.P. Arovas and A. Yacoby for
enlightening discussions, A. Geim for supplying the natural graphite
and M. Brongersma, A. Guichard and E. Barnard for the Raman
spectrometry. This work was supported by MARCO/FENA program and the
Office of Naval Research. K.T. was supported by a Hertz Foundation
graduate fellowship, N.S was supported by a William R. and Sara Hart
Kimball Stanford Graduate Fellowship, J.A.S. by a National Science
Foundation graduate fellowship. B.H. was supported in part by the
Lavoisier fellowship. Work was performed in part at the Stanford
Nanofabrication Facility of NNIN supported by the National Science
Foundation under Grant ECS-9731293.

\bibliographystyle{apsrev}

\end{document}